\documentclass[10pt,twocolumn,letterpaper]{article}

\usepackage[pagenumbers]{iccv} % To force page numbers, e.g. for an arXiv version

\definecolor{iccvblue}{rgb}{0.21,0.49,0.74}
\usepackage[pagebackref,breaklinks,colorlinks,allcolors=iccvblue]{hyperref}
\usepackage{bm}
\usepackage{multirow}
\usepackage{graphicx}
\usepackage{amsmath}
\usepackage{amssymb}

\title{S2MNet: Speckle-To-Mesh Net for Three-Dimensional Cardiac Morphology Reconstruction via Echocardiogram}

\author{
\textbf{Xilin Gong$^{1,*}$, Yongkai Chen$^{2, *}$, Shushan Wu$^{1}$, Fang Wang$^{3}$}\\
\textbf{Ping Ma$^{1,\dagger}$, Wenxuan Zhong$^{1,\dagger}$}\\
$^1$ Department of Statistics, University of Georgia \quad\\ $^2$Department of Statistics, Havard University \quad \\ $^3$Institute of Geriatric Medicine, Chinese Academy 
of Medical Sciences, Beijing Hospital  \\ 
}
\begin{document}

\maketitle

\begin{abstract}

Echocardiogram is the most commonly used imaging modality in cardiac assessment duo to its non-invasive nature, real-time capability, and cost-effectiveness. Despite its advantages, most clinical echocardiograms provide only two-dimensional views, limiting the ability to fully assess cardiac anatomy and function in three dimensions. While three-dimensional echocardiography exists, it often suffers from reduced resolution, limited availability, and higher acquisition costs. To overcome these challenges, we propose a deep learning framework S2MNet that reconstructs continuous and high-fidelity 3D heart models by integrating six slices of routinely acquired 2D echocardiogram views. Our method has three advantages. First, our method 
avoid the difficulties on training data acquasition by simulate six of 2D echocardiogram images from corresponding slices of a given 3D heart mesh. 
Second, we introduce a deformation field-based method, which avoid spatial discontinuities or structural artifacts in 3D echocardiogram reconstructions. We validate our method using clinically collected echocardiogram and demonstrate that our estimated 
left ventricular volume, a key clinical indicator of cardiac function, is strongly correlated with the doctor measured GLPS, a clinical measurement that should demonstrate a negative correlation with LVE in medical theory. This association confirms the reliability of our proposed 3D construction method.


\end{abstract}

% \IEEEoverridecommandlockouts
% \begin{keywords} 
% 3D Reconstruction, Echocardiogram, Medical Image Process, Healthcare
% \end{keywords}

% \IEEEpeerreviewmaketitle

\section{Introduction}
%\subsection{Motivation and background}
Echocardiogram (echo) has become an indispensable tool in cardiac care, widely used for its non-invasive nature, accessibility, and ability to provide high spatial and temporal resolution imaging of cardiac structures \cite{article, nature1}. In echocardiograms, cardiac tissues are represented as speckles, which facilitate the visualization of the morphology and enable real-time tracking of both local and global cardiac movements for functional assessment \cite{RN11}. Moreover, recent advances in machine learning have enhanced the fast and accurate automated diagnosis of echocardiograms for conditions such as valvular abnormalities, and coronary artery disease \cite{zhangjecho}.

%Clinicians rely on echocardiogram to provide immediate insights into the function of the heart \cite{zhangjecho}, facilitating the rapid diagnosis of conditions such as heart failure, valvular abnormalities, and congenital defects.

For comprehensive cardiac evaluation and benchmarking the automated diagnostics, three-dimensional visualization is the ideal solution \cite{Houck2006Live3E}.
However, most echocardiography systems are limited to providing multiple two-dimensional (2D) views \cite{gillam2024echocardiography}.
This situation persists because 2D echo is more accessible, cost-effective, and has higher spatial and temporal resolution. In comparison, 3D echo often suffers from lower resolution, in addition to higher costs in the equipment and limited accessibility \cite{turton2017role, 2dlimit}.

Deep neural network has emerged as a powerful computational approach for reconstructing 3D objects by using 2D images \cite{2dto3d}.
However, unlike the traditional 3D reconstruction task where the images are taken from multiple views to capture the surface details,
the echocardiogram directly measures structural depths within its scanning plane. 
Therefore, general-purpose 3D reconstruction deep learning models (e.g., Pix2Vox \cite{pix2vox}) and the associated training datasets are not applicable to our task.
%usually only taken from few standard views %\cite{henry2022three}. 
What's more, training a specialized deep learning model for 3D cardiac morphology reconstruction faces two primary challenges: learning difficulties and limited annotated datasets.
Since echocardiogram is acquired via 2D scanning planes, it inherently lacks complete spatial information. This limitation is compounded by variability in scanning angles, patient motion, and operators' experience. %—all of which degrade reconstruction reliability.
Another critical bottleneck is the lack of high-quality 2D echocardiogram images paired with accurate 3D ground-truth annotations \cite{e-p2vforheart}, which prevent implementing large-scale supervised learning.

In this article, we propose a novel framework designed to address the challenges of 3D heart reconstruction from limited echocardiogram views. By integrating a series of artificial intelligence approaches, our method effectively reconstructs continuous and high-fidelity heart models.
We have the following contributions.

\begin{enumerate}
\item We propose a deep learning framework that reconstructs high-precision 3D heart models from multiple 2D echocardiogram views. Leveraging CycleGAN to bridge the domain gap, synthetic 2D training data is translated into the echocardiogram domain, making it closely resemble real clinical data. This allows the model to generalize effectively and be directly applicable to real-world cases without additional fine-tuning.

\item We address the discontinuities in predictive models by replacing binary voxel representation with a deformation field-based approach. Traditional voxel-based prediction methods divide space into discrete grids, often causing in defects and holes. In contrast, our method predicts a continuous vector field, which, when applied to a template, ensures smooth and accurate 3D reconstruction with continuous surfaces, enhancing structural integrity in cardiac modeling.

\item We validate our method on real echocardiogram data from clinical cases, demonstrating its effectiveness in real-world medical settings. By reconstructing patient-specific 3D meshes and calculate the volume of the left ventricle, we enable accurate ejection fraction (EF) computation. The strong correlation between EF and global longitudinal peak strain (GLPS), with a Pearson correlation coefficient of -0.82, confirms the clinical reliability of our reconstruction.
\end{enumerate}

This article is organized as follows. In Section \ref{sec:related}, we reviewed related works about 3D reconstruction on general and ultrasound images. In section \ref{sec:dataset}, we introduced the datasets we used in this study. We present our proposed framework stage by stage in section \ref{sec:methods} and present the results and analysis in section \ref{sec:result}. Finally, we conclude this article with limitations and future work of this article in section \ref{sec:discussion}.

\section{Related works} \label{sec:related}
Current 3D reconstruction algorithms fall into two main categories. The first one focuses on general images taken in natural environments and makes up the majority of existing methods. The second one, which is less explored, targets medical images despite their promising applications in healthcare.

\subsection{3D Reconstruction Algorithms for General Images}
Generally speaking, 3D reconstruction for general images includes two categories: geometric methods that rely on camera or depth data, and deep learning methods like encoder-decoders and NeRF, which require ground-truth 3D or known camera views. Both face challenges when applied to echocardiograms.
\subsubsection{Geometric relationship-based method}

Existing 3D reconstruction approaches, extensively developed for non-medical imaging, highlight both the potential and the limitations of current technology when applied to ultrasound. Traditional techniques for 3D image reconstruction in natural images often leverage geometric relationships, such as those used in Structure from Motion (SfM) \cite{SFM1, SFM2}, Multi-View Stereo (MVS) \cite{msv1, msv2, msv3}, Kinect Fusion \cite{KinectFusion1, KinectFusion2}, and Point Cloud Registration \cite{PCR1, PCR2, PCR3}. These methods rely on geometric relationships to minimize the reprojection error \cite{hartley2004multiple}, and often do not require explicit ground-truth 3D data, which has led to successful applications in various fields. However, these approaches also depend on inputs like camera positions, RGB-D images, or point cloud data, which are generally unavailable or impractical to obtain in the context of echocardiogram. As a result, direct application of these methods to 3D cardiac reconstruction is unsuitable, necessitating new approaches tailored to the complexities of echocardiogram imaging.

\subsubsection{Deep learning-based method}

In recent years, deep learning-based methods have emerged as powerful tools for 3D reconstruction from 2D images, offering substantial advantages over geometric relationship-based algorithms. These approaches can be broadly divided into two main categories: encoder-decoder frameworks \cite{encoder1, encoder2, pointoutnet, popov2020corenet, pix2vox} and Neural Radiance Fields (NeRF) \cite{nerf++, nerf1, dnerf}. 

%\begin{enumerate}

Encoder-decoder methods typically use convolutional neural networks (CNNs) to transform input images into latent representations, which are then decoded into 3D reconstructions \cite{dl3dsurvey}. These approaches have demonstrated improved performance over geometric methods, especially in reconstructing intricate details. However, they often require ground-truth 3D models for training, which is a significant limitation in many medical applications. Additionally, encoder-decoder networks may produce outputs with low resolution and discontinuities, which can impact the precision of the reconstructed model.

NeRF represents an alternative, leveraging volume rendering to produce continuous, detailed reconstructions by modeling the scene as a 3D field with a multilayer perceptron (MLP). Enhanced by positional encoding, the MLP learns geometry and appearance directly from image supervision, enabling smoother and more accurate reconstructions than traditional methods. However, NeRF requires 2D RGB images with known camera poses \cite{nerfreview}, which are unavailable in uncontrolled echocardiogram settings.

%\end{enumerate}

\subsection{3D Reconstruction Algorithms for Ultrasound Images}

Adapting 3D reconstruction methods for ultrasound images presents additional challenges due to the nature of ultrasound data and the complex anatomy of structures such as the heart. Previous approaches have been attempted in areas other than echocardiogram. For example, \cite{3Drebifur} used Canny's edge detection method \cite{canny1986computational} on ultrasound images of the carotid bifurcation, where multiple images were acquired in parallel along the artery and then stitched together into a 3D model based on detected edges. However, this method is impractical for echocardiogram due to the heart’s intricate structure and movement, which reduces the effectiveness of simple edge-detection techniques.

Deep learning has been explored for ultrasound-based 3D reconstruction with some success in other anatomical structures. For instance, \cite{3Dresnet} applied a 3D-ResNet model to reconstruct high-quality 3D images of the fingers, radial and ulnar bones, and metacarpophalangeal joints. However, this method has not yet been applied to the heart. Another notable study \cite{e-p2vforheart} employed an encoder-decoder CNN framework to reconstruct 3D cardiac models from 2D echocardiogram by using a voxel-based approach. However, the reconstructions often show instability, with holes and missing regions in real echocardiogram images, limiting clinical usefulness.
% However, the reconstructed models often exhibit instability, with frequent issues such as holes and incomplete regions when applied to unseen 2D images, which reduces the clinical utility of the reconstructions.

\section{Datasets} \label{sec:dataset}
Two datasets are used in this study: a published 3D heart mesh dataset \cite{3Ddataset} and a real-world echocardiogram dataset collected by us from clinical patients. The former is used to generate synthetic training images and provide supervision for training S2MNet, while the latter is used to validate our method and demonstrate its potential for clinical diagnosis.

\subsection{3D Heart Mesh Dataset}
This dataset is collected from end-diastolic CT scans of patients who visited the emergency room with acute chest pain but were ultimately diagnosed with asymptomatic hearts \cite{3Ddataset}. The dataset comprises 1,000 four-chamber heart meshes in point cloud format, each annotated with 24 tissue labels indicating different heart regions. In our framework, it is used to generate 6 views 2D synthetic training images. 

\subsection{Real Dataset from clinical patients}
This dataset is collected from patients with heart failure at Beijing Hospital. The dataset comprises six-angle echocardiogram images of 14 heart failure patients. For each patient, the GLPS value provided by a physician serves as a clinical reference to validate our framework. The study was approved by Beijing Hospital Ethics Committee (1100000185432).

\section{Methods} \label{sec:methods}
Our framework for reconstructing a high-quality, continuous 3D cardiac model from 2D echocardiogram images consists of four primary stages.

\begin{figure*}[h]
\centering
\includegraphics[width=0.8 \linewidth]{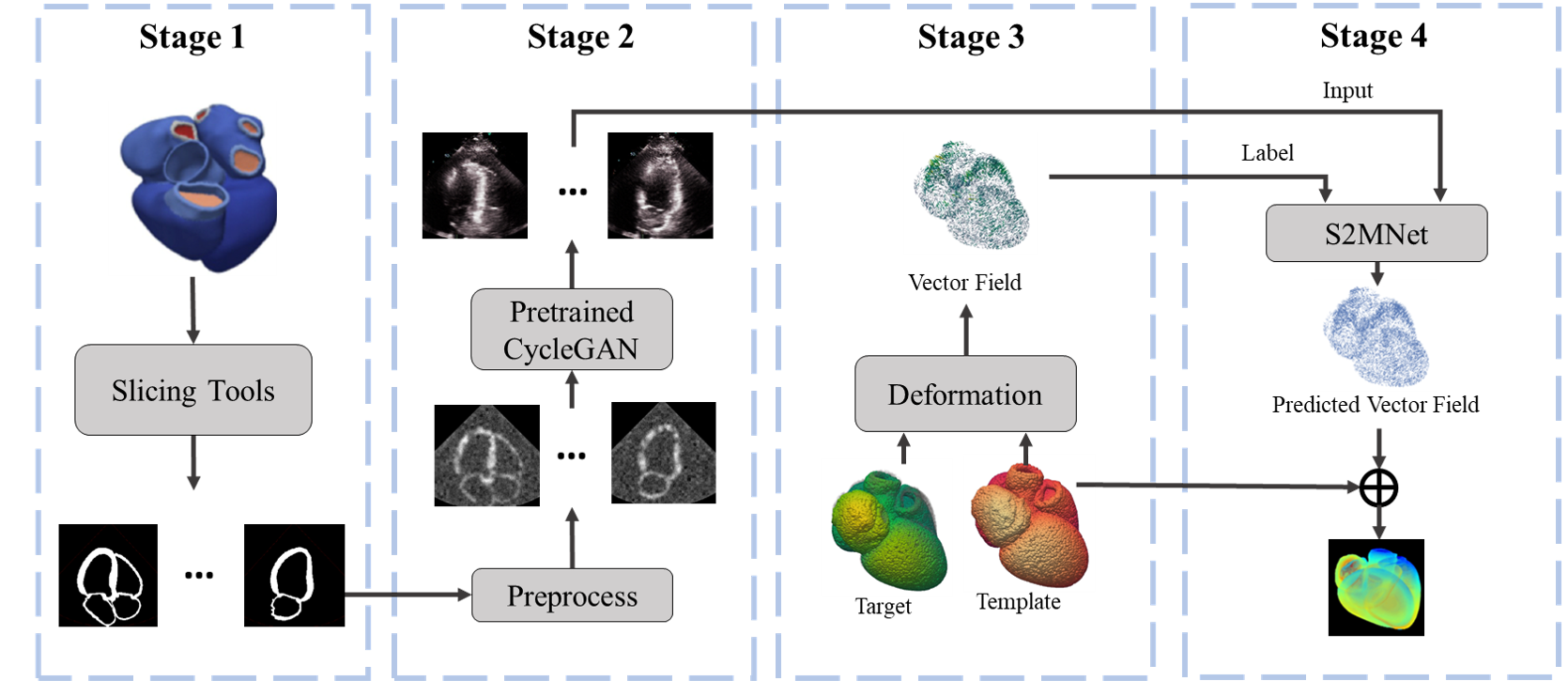}
\caption{An overview of our 3D cardiac morphology reconstruction framework. In stage 1, synthetic 2D cardiac image slices are generated from a 3D heart model. In stage 2, these binary images are preprocessed and passed through a pretrained CycleGAN to their similarity to real echocardiogram images. In stage 3, a vector field is computed between a template heart mesh and other cardiac meshes through deformation method. In stage 4, S2MNet is trained to predict vector fields from ultrasound-styled images, which are added to the template mesh to reconstruct the 3D cardiac structure. }\label{framework}
\end{figure*}

An overview of our proposed framework, detailing each stage, is illustrated in Figure \ref{framework}. Through this structured approach, our framework achieves precise 3D cardiac modeling that reflects real-world complexities of cardiac anatomy and motion. We will present our framework stage by stage.

\subsection{Slicing 2D Images from 3D Cardiac Models}
Acquiring real-time echocardiogram images alongside 3D cardiac models is often impractical in clinical settings due to equipment and procedural constraints. To overcome this limitation, we leveraged a dataset of 1,000 four-chamber heart models derived from end-diastolic CT scans, as described in \cite{3Ddataset}. These models serve as a basis for generating synthetic training data in 2D, closely emulating the scanning perspectives typically employed in clinical practice.

To achieve realistic and clinically relevant 2D image slices, we developed a simulation software using PyVista. This software replicates the scanning methodology utilized by medical professionals, allowing us to create slices from the 3D models in a way that better matches the orientations and spatial relations observed in real-world echocardiogram imaging. Unlike straightforward slicing techniques that cut along predefined planes, our software aligns the cardiac model within a body mesh for a more holistic, realistic representation of imaging conditions.

The first step in our approach involves positioning the cardiac model within a human body mesh, as provided by \cite{bodymesh}, to simulate a realistic anatomical context. We develop a software tool that enables interactive manipulation of the heart model within the 3D mesh through translation, rotation, and scaling. The heart model is represented as a set of points defined by their 3D coordinates $ \mathbf{p} = [x,y,z]^T$. Translation is performed by adding a displacement vector 
$\mathbf{\Delta}=[\Delta x, \Delta y, \Delta z]^T$, resulting in the transformed point: $\mathbf{p'}=\mathbf{p}+\mathbf{\Delta}$. Rotation is implemented using standard 3D rotation matrices $R \in \mathbb{R}^{3 \times 3}$, which can be composed from individual rotations around the x-, y-, and z-axes. The rotated point is computed as $\mathbf{p}'=R\mathbf{p}$, where $R=R_z(\theta_z)R_y(\theta_y)R_x(\theta_x)$. The rotation matrix $R_x(\theta_x)$, $R_y(\theta_y)$ and $R_z(\theta_z)$ around x-, y-, and z-axes is defined in \cite{lengyel2011mathematics}. 
Scaling is achieved by multiplying each coordinate by a scalar $s \in \mathbb{R}$, resulting in $\mathbf{p}'=s\mathbf{p}$. These transformation methods consistent with common practices in 3D computer graphics \cite{3dgraph}. We control the heart mesh by specifying the input parameters $\Delta x, \Delta y,\Delta z$ for translation, $ \theta_x, \theta_y, \theta_z$ for rotation and $s$ for scaling.
These adjustments simulate anatomical variability across patients and provide a robust foundation for generating diverse synthetic data.

% The rotation matrix around the z-axis by angle $\theta_z$ is defined in Equation \ref{eq:rotz}, while the rotations around the
% x- and y-axes are defined in Equations \ref{eq:rotx} and \ref{eq:roty}.
% \begin{equation}
%     R_z(\theta) = 
% \begin{bmatrix}
% \cos\theta & -\sin\theta & 0 \\
% \sin\theta & \cos\theta & 0 \\
% 0 & 0 & 1
% \end{bmatrix}
% \label{eq:rotz}
% \end{equation}

% \begin{equation}
%     R_x(\theta) = 
% \begin{bmatrix}
% 1 & 0 & 0 \\
% 0 & \cos\theta & -\sin\theta \\
% 0 & \sin\theta & \cos\theta
% \end{bmatrix}
% \label{eq:rotx}
% \end{equation}

% \begin{equation}
% R_y(\theta) = 
% \begin{bmatrix}
% \cos\theta & 0 & \sin\theta \\
% 0 & 1 & 0 \\
% -\sin\theta & 0 & \cos\theta
% \end{bmatrix}
% \label{eq:roty}
% \end{equation}
% \textcolor{blue}{Our software is available at...  I can publish the code of it on Github or it is not necessary to do that?}

The second step involves generating standardized slices in clinically relevant views. Six standard cardiac imaging views are simulated to reflect the orientations and positions used in echocardiogram examinations. Specifically, three long-axis views (Apical 4-chamber, Apical 2-chamber, and Parasternal long-axis) and three short-axis views (Basal, Mid-cavity, and Apical) are generated. For the long-axis views, the software simulates the orientation of the ultrasound probe by positioning the virtual sector origin at a fixed point while varying the sector axis angle to achieve the specific viewing angles typical of each long-axis slice. For each short-axis view, the software generates slices by aligning the sector at different axial depths relative to the heart.

This standardized slicing framework ensures that the synthetic data are aligned with clinical practice, enabling our model to be trained on a range of realistic perspectives that mirror the scanning angles and orientations used by clinicians. The specific coordinates and landmarks for each simulated view are provided in Table \ref{tab:6view}, detailing the parameters used in our synthetic dataset generation process.

\begin{table*}
  \begin{center}
    \caption{Echocardiogram views with their corresponding anatomical landmarks and coordinates used to define each sector. These sector definitions were suggested by a clinical expert to ensure anatomical accuracy.}\label{tab:6view}
    \vspace{5pt}
    \begin{tabular}{c| c| c}
    \hline
      View&  Landmarks&Coordinates \\
      \hline
Apical 4-chamber& Left Atrium, Right Atrium, Left Ventricle Apex & (62.63, -60.94, -28.13)\\
Apical 2-chamber & Left Ventricle Apex, Mitral Valve, Right Ventricle & (62.63, -60.94, -28.13)  \\
Apical 3-chamber & Left Ventricle Apex, Mitral Valve, Aortic Valve&(62.63, -60.94, -28.13)\\
Basal&Left Ventricle Apex, Whole Heart&(16.27, -8.42, -9.61)\\
Mid-cavity&Left Ventricle Apex, Whole Heart&  (39.04, -20.21, -23.07)\\
Apical & Left Ventricle Apex, Whole Heart & (45.55, -23.58, -26.91)\\
\hline
\end{tabular}
\end{center}
% \vspace{-20pt}
\end{table*}

\subsection{Echocardiogram Style Transfer Using CycleGAN}

To create realistic echocardiogram-style training data, we use a two-step approach for image style transfer: 1) preprocessing synthetic images by cropping and adding Gaussian noise, and 2) applying a CycleGAN model \cite{cyclegan} for translating these synthetic images into the echocardiogram style.

First, we preprocess the images by cropping and adding Gaussian noise following the settings in \cite{e-p2vforheart}. Each synthetic image is cropped using a sector mask that mimics the shape and coverage of a typical ultrasound scan. Within this sector, the heart region is set to white, while the background is black, representing areas where ultrasound waves would not detect tissue. Gaussian blurring and random noise are then applied within this mask to replicate the granular texture characteristic of ultrasound images. This preprocessing step generates “pseudo” images that approximate ultrasound scans, though without true ultrasound artifacts and variances.

Second, we perform style transfer to echocardiogram using CycleGAN. Traditional image-to-image translation relies on paired training images to learn a mapping from a source to a target domain. However, obtaining paired real echocardiogram that exactly correspond to our synthetic cardiac slices is impractical. To address this, we use CycleGAN, which enables translation from the pseudo-synthetic domain $X$ to the echocardiogram domain $Y$ without requiring paired samples.

Two mapping functions $ G : X\rightarrow Y$,  $F: Y\rightarrow X$ and their associated adversarial discriminators $D_Y$ and $D_X$ are trainable in this model. An adversarial loss is applied to trains the generator to produce images indistinguishable from real images in the target domain by fooling a discriminator. The adversarial loss for mapping function $F$ and discriminator $D_Y$ is:
\begin{equation}
\mathcal{L}_{GAN}(G, D_Y) = \mathbb{E}_{y}[\log D_Y(y)] + \mathbb{E}_{x}[\log(1 - D_Y(G(x)))]
\end{equation} 
The discriminator outputs a value between 0 and 1, representing the probability that an input image is a real sample from the target domain. $G$ aims to minimize this loss function against $D_Y$ tries to maximize it, i.e, $min_G max_{D_Y}\mathcal{L}_{GAN}$. The adversarial loss for mapping function $G$ and discriminator $D_X$ is similar. In order to enforce cycle consistency, ensuring that an image translated to the other domain and back returns to the original, a cycle consistency loss is introduced: 

\begin{equation}
 \mathcal{L}_{cycle}(G, F) = \mathbb{E}_{x}[\|F(G(x)) - x\|_1] + \mathbb{E}_{y}[\|G(F(y)) - y\|_1]
\end{equation}

The full loss function is
\begin{equation}
    \begin{split}
            \mathcal{L}(G,F, D_X, D_Y) &=  \mathcal{L}_{GAN}(G,D_Y,X,Y) \\ &+\mathcal{L}_{GAN}(F,D_X,X,Y)\\
    &+\lambda \mathcal{L}_{cycle}(G,F)
    \end{split}
\end{equation}

During training, the CycleGAN model captures the essential texture, noise patterns, and grayscale contrasts specific to echocardiogram, allowing us to generate realistic echo-like images from the pseudo-synthetic inputs.

Once these echo-style images are generated, they are ready for use in training the 3D reconstruction model, providing it with data that closely resembles actual echocardiogram scans.

\subsection{Cardiac Deformation Field-Based Labeling}
The cardiac deformation field is what we want to predict from our S2MNet, by directly add it to reference heart mesh, we can get continuous heart surface.
Given that the vertices of cardiac meshes in our training dataset lack direct one-to-one correspondence, we establish consistent anatomical correspondences by computing cardiac deformation fields between a reference heart mesh and all other meshes.
This alignment not only standardizes the learning task for cardiac mesh reconstruction but also enables automatic annotation of anatomical labels from the reference mesh to predicted meshes, ensuring consistent vertex-wise annotation across the dataset.

To accurately compute the cardiac deformation effectively capturing both subtle and significant morphological differences between the heart meshes,  
we calculate the deformation field with the robust optimal transport method \cite{nipsot}.
Suppose we have a reference heart mesh consisting of $N$ vertices $\{p_i\}_{i=1}^N$ and a target heart mesh consisting of $M$ vertices $ \{q_j\}_{j=1}^M$, where $p_i,q_j \in  R^3$.
We first solve a weighted assignment between the vertices of the reference mesh  and the target mesh by
\begin{equation}
    \begin{split}
           argmin_{\Pi = [\pi_{i,j}]}    & 
            \sum_{i=1}^{N} \sum_{j=1}^{M}  \frac{\pi_{i,j}}{2} \| p_i - q_j \|^2_2 + \tau^2 KL \left(P_{\Pi} \textbf{1}_M || P_\alpha\right)  \\
         &  + \tau^2 KL \left(  P_{\Pi}^\prime \textbf{1}_N || \beta_j\right)+ \sigma^2 KL (P_{\Pi} || \alpha_i \otimes \beta_j),
    \end{split}
\end{equation}
where $\Pi= [\pi_{i,j}]$ is the probability assignment matrix with $\pi_{i,j}  \geq 0$ and $_{i=1}^{N} \sum_{j=1}^{M} \pi_{i,j}  = 0$.
$KL(\cdot \|\cdot)$ is the Kullback–Leibler divergence.
$P_{\Pi}$ is the joint probability mass function of two discrete random variables with the probability matrix as $\Pi$.
$\tau$ is the tuning parameter that controls the changes in the vertice weights after assignment.
$\sigma$ is the tuning parameter that regularizes the assignment, encouraging it to be more uniform or smoother. The values of 
$\tau$ and $\sigma$ are selected according to the default settings used in \cite{nipsot}.

The displacement for aligning the reference heart mesh with the target for the point $p_i$ is calculated as
$v_i=\frac{\sum_{j=1}^{\mathrm{M}} \pi_{i, j} \cdot\left(q_j-p_i\right)}{\sum_{j=1}^{\mathrm{M}} \pi_{i, j}} \in \mathbb{R}^3.$
Given the results $\{p_i,v_i\}_{i=1}^N$, the deformation field for the reference heart mesh is learned through training a PointPWC-Net  \cite{wu2019pointpwc} and fine-tuned with the spline estimator using Gaussian kernel. 

% A cardiac deformation field describes the spatial transformation needed to map one heart structure onto another, effectively capturing both subtle and significant morphological differences between subjects. 
% For example, if we have a standard template of a heart structure, the deformation field for a specific subject would show how each point in the template needs to shift to match the subject's unique heart structure.

% \textcolor{blue}{We calculate the deformation field with the robust optimal transport method \cite{balaji2020robust, nipsot}. Instead of nearest-neighbor projections or chamfer distance relied traditional approaches, this method optimizes transport plans using a soft probability-based OT approach. This problem can be formulated as: $
% \text{OT}_{\sigma,\tau}(A,B) = \min_{\pi_{i,j} \geq 0} \sum_{i=1}^{N} \sum_{j=1}^{M} \pi_{i,j} \cdot \frac{1}{2} \| p_i - q_j \|^2_{R^D}
% + \sigma^2 KL (\pi_{i,j} || \alpha_i \otimes \beta_j) + \tau^2 KL \left(\sum_j \pi_{i,j} || \alpha_i\right) + \tau^2 KL \left(\sum_i \pi_{i,j} || \beta_j\right) $, where $KL(a || b)$ denotes the Kullback-Leibler divergence to enforce soft constraints, $\sigma$ is a parameter that controls the blur radius for soft assignments and $\tau$ controls the maximum reach distance for valid correspond points. After a simple OT-based prealignment method, a deep neural network is applied to predict deformations. }

%To ensure continuous cardiac morphology learning, we use a technique called deformation field analysis. 
 
\subsection{S2MNet for 3D reconstruction}

% \textbf{If we use a variational autoencoder, could it be helpful for making uncertain quantifications?}

\begin{figure*}[h]
\centering
\includegraphics[width=0.8\linewidth]{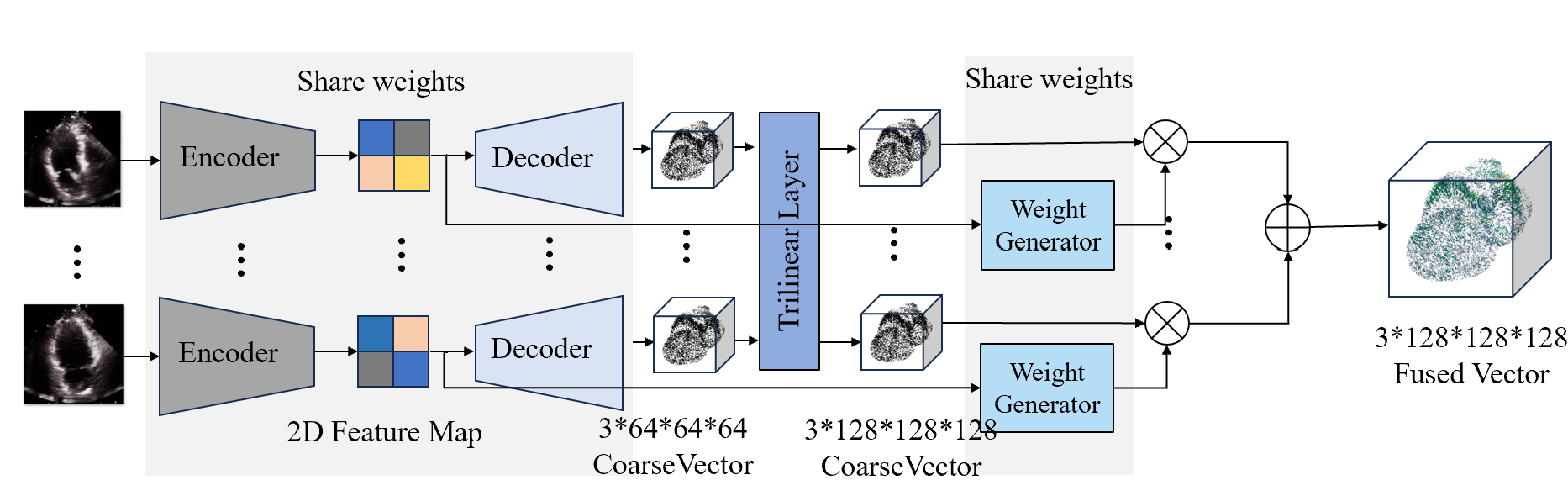}
\caption{An overview of our S2MNet for 3D reconstruction. The encoder extracts 2D features, while the decoder and trillinear layer generates coarse vector fields. A weight generator assigns weights to each coarse vector field, which are fused via weighted summation to produce the final vector field. The output is a deformation vector which should be added to the template to get 3D heart mesh.}\label{network}
\end{figure*}

% \begin{center}
% \includegraphics[width=0.3\linewidth]{figures/distancewg.png}

% \figurecaption{weight generation module}\label{distancewg}
% \end{center}

For 3D reconstruction, we build S2MNet which allows us to recover a 3D model of the object from any number of uncalibrated 2D images. An overview of our S2MNet is shown in Figure \ref{network}. The encoder consists of a ResNet50 \cite{resnet} backbone followed by three 2D convolutional layers and one 3D convolutional layer. The decoder is composed of six transposed 3D convolutional layers, followed by a trilinear upsampling layer to achieve higher resolution. The encoder output is also passed to a weight generator, which includes six 3D convolutional layers and produces weights for each coarse vector field generated by the decoder. Finally, each coarse vector field is multiplied by its corresponding weight, and the weighted fields are summed to obtain the final fused vector field. Compared with the widely used voxel-based methods \cite{pix2vox, e-p2vforheart}, there are two differences which make our method more suitable for heart reconstruction task. First, to ensure the reconstructed object has a continuous surface, our model outputs a deformation field rather than a binary voxel array. While voxel-based approach enables 3D reconstruction, it often results in noticeable defects and holes. Our method instead predicts a deformation vector field, which is added to a template to produce the continuous 3D object. Second, improve efficiency, we use a trilinear interpolation layer for higher resolution. Benifit from the deformation field, our method easily produces high-res 3D objects with minimal overhead. In contrast, voxel-based methods require scaling the entire network, increasing time and memory costs, which limits practical use in medical settings.

\section{Results and Analysis}\label{sec:result}

In section \ref{sec:cyclegan}, we present the result of the first and second stage of our framework, which are used to generate synthetic training data. In section \ref{sec:s2m}, we present the 3D reconstruction result of the last stage of our framework. Then in section \ref{sec:segmentation} and \ref{sec:real}, we valid our 3D reconstruction result and analysis it on real clinical data.

\subsection{Generated Synthetic Data from Slicing and Style Transfer} \label{sec:cyclegan}
We generate realistic echocardiogram-style training data using a two-step image style transfer approach. First, synthetic images are preprocessed by cropping within a sector mask and adding Gaussian noise to mimic echocardiogram texture. Then, a CycleGAN model is employed to translate these pseudo-synthetic images into the echocardiogram domain. The data generated results of each step are shown in Table \ref{tab:image_table}.

\begin{table*}[h]
    \centering
    \caption{The images generated at each stage of synthetic echocardiogram image creation. From up to down: Binary segmentation mask from 3D model generated by the slicing stage, Pseudo image with noise generated the preprocessing of CyeleGAN, Synthetic echocardiogram style image generated by CycleGAN.}
    \begin{tabular}{c|cccccc}  
        \hline
        & \multicolumn{6}{c}{View} \\ 
        & A4C & A2C & A3C & Basal & Mid\_cavity & Apical \\ \hline
        Binary
        & \includegraphics[width=0.1\linewidth]{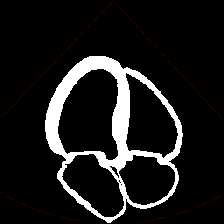} 
        & \includegraphics[width=0.1\linewidth]{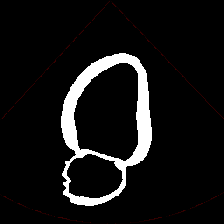} 
        & \includegraphics[width=0.1\linewidth]{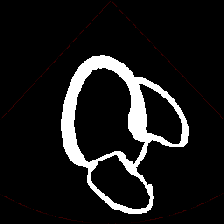} 
        & \includegraphics[width=0.1\linewidth]{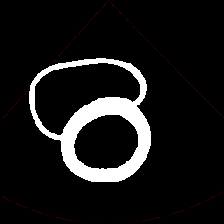} 
        & \includegraphics[width=0.1\linewidth]{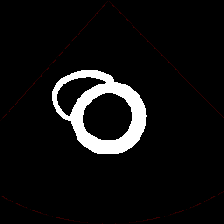} 
        & \includegraphics[width=0.1\linewidth]{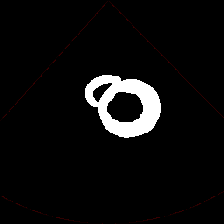} \\
        Pseudo
        & \includegraphics[width=0.1\linewidth]{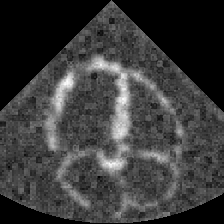} 
        & \includegraphics[width=0.1\linewidth]{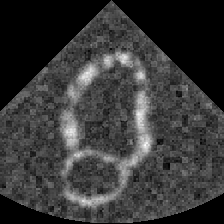} 
        & \includegraphics[width=0.1\linewidth]{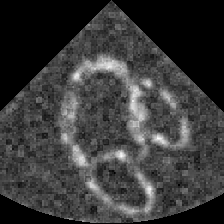} 
        & \includegraphics[width=0.1\linewidth]{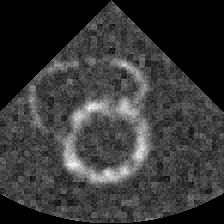} 
        & \includegraphics[width=0.1\linewidth]{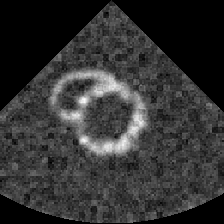} 
        & \includegraphics[width=0.1\linewidth]{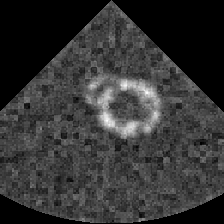} \\ 
        Echocardiogram
        & \includegraphics[width=0.1\linewidth]{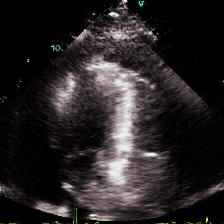} 
        & \includegraphics[width=0.1\linewidth]{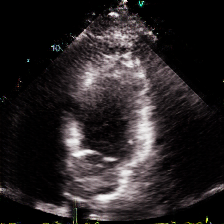} 
        & \includegraphics[width=0.1\linewidth]{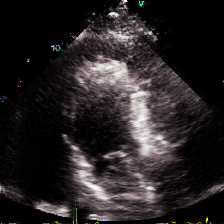} 
        & \includegraphics[width=0.1\linewidth]{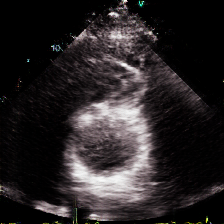} 
        & \includegraphics[width=0.1\linewidth]{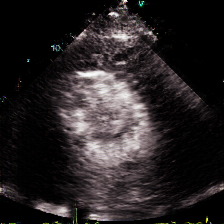} 
        & \includegraphics[width=0.1\linewidth]{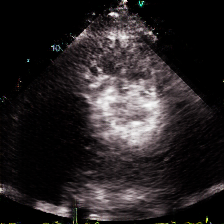} \\ 
        \hline
    \end{tabular}
    \label{tab:image_table}
\end{table*}

\subsection{3D Reconstruction Result of S2MNet}\label{sec:s2m}

% \begin{table}
% \captionsetup{width=\linewidth}

% \begin{tabular}{c|ccc}
% \multirow{2}{*}{Methods} & \multicolumn{3}{c}{CycleGAN inferenced} \\ 
%  & 6 views       & A2C         & A4C        \\ \hline
% PiVox/Fast               &   0.661 & 0.644      & 0.652      \\
% E-PiVox/Fast             &   0.667  & 0.651       & 0.660      \\
% PiVox/Accurate          &   0.738   & 0.673       & 0.732      \\
% E-PiVox/Accurate         &   0.749   & 0.674       & 0.734      \\
% Pi2Mesh (Ours)          &   $0.775 \pm 0.0123$ &   0.702      &  0.744    \\   
% \end{tabular}
% \label{tab:result_of_iou}
% \end{table}

\begin{figure}[h]
\centering
\includegraphics[width=0.6\linewidth]{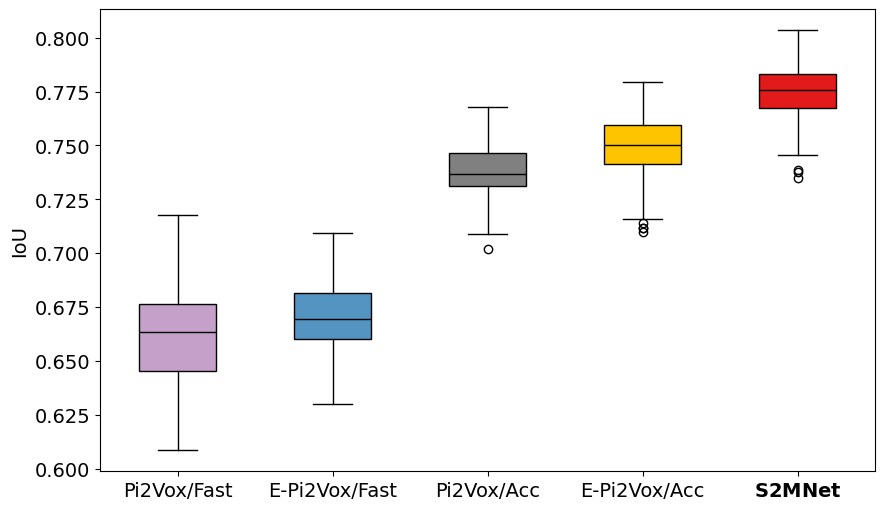}
\caption{A box-and-whisker plot for IoU comparison of different reconstruction methods at a resolution of $128^3$. Pi2Vox and E-Pi2Vox refer to methods proposed in \cite{pix2vox, e-p2vforheart}. "Fast" and "Acc" indicate different network architectures, with "Fast" using fewer layers for faster inference. Our method is shown in red on the left.}
\label{fig:result_of_iou}
\end{figure}

%For the synthetic dataset, there are 1000 3D meshes in total. 
We randomly split the synthetic dataset into a training dataset and testing dataset with ratio of 80\% versus 20\%. Notably, the testing dataset was exclusively reserved for the evaluation phase and was not utilized during the training processes of CycleGAN and S2MNet.

Quantitatively assessing our results, we employed the mean squared error (MSE) as the evaluation metric, yielding a remarkably low MSE of $1.84\times10^{-2}$ with standard deviation $7.71\times10^{-3}$. This low error value underscores the high accuracy achieved by our 3D reconstruction method. As a baseline, random guesses from a uniform distribution over the target range yielded a much higher MSE of $2.77\times10^1$ with variance $1.57\times10^{1}$. To ensure a fair comparison with other methods \cite{pix2vox, e-p2vforheart} that perform 3D echocardiogram reconstruction in voxel format, we convert the point-based outputs into voxel representations using their spatial coordinates, then evaluate the overlap between predicted and ground-truth voxels using the Intersection over Union (IoU) metric \cite{iou}. The IoU is defined as: $IoU=\frac{TP}{TP+FP+FN}$, where TP (true positives) indicates correctly predicted occupied voxels, FP (false positives) are predicted occupied voxels that are not in the ground truth, and FN (false negatives) are ground-truth occupied voxels missed by the prediction. The results, presented in Figure \ref{fig:result_of_iou}, show that our method achieves state-of-the-art performance.

\begin{figure*}[h]
\centering
\includegraphics[width=0.6\linewidth]{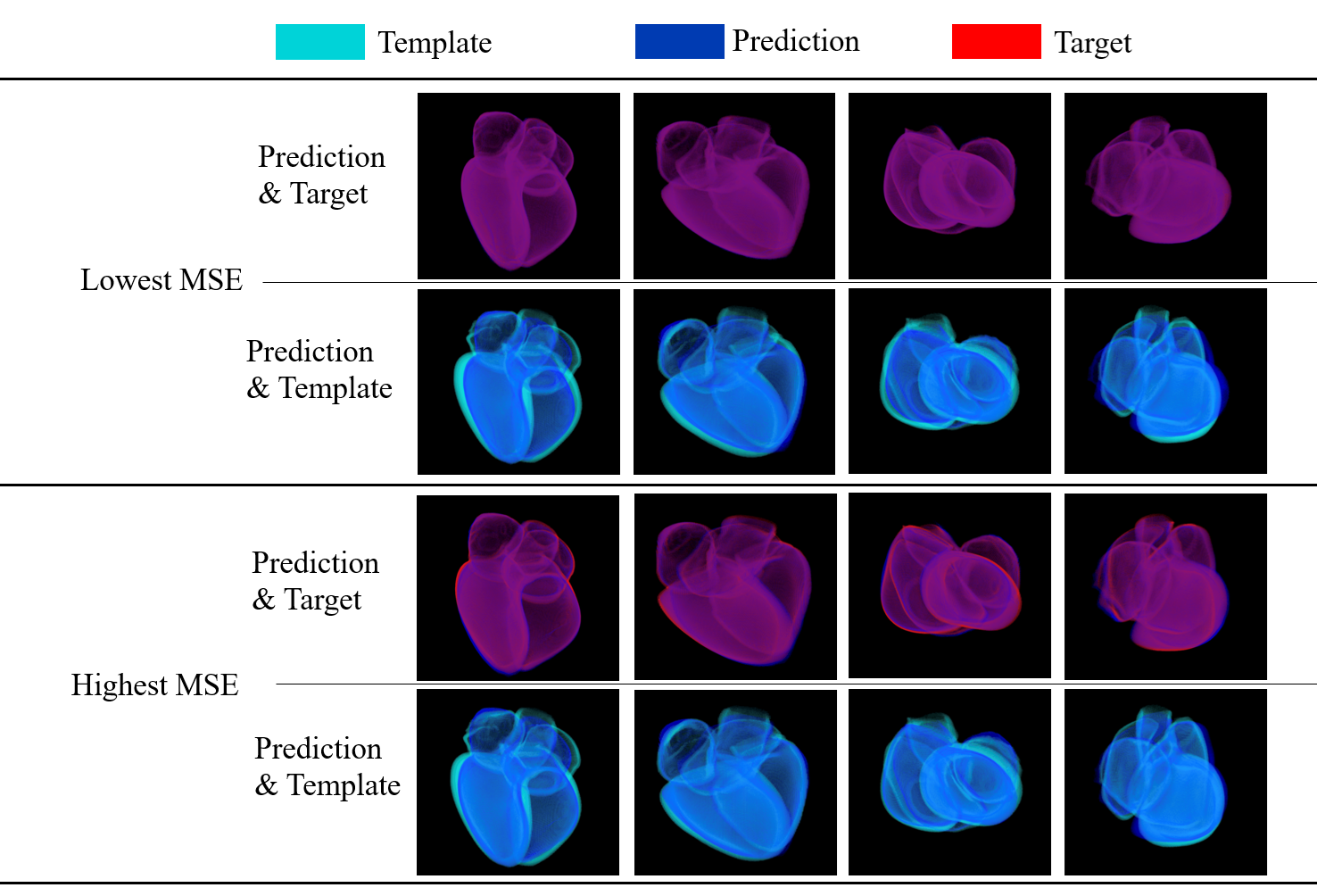}
\caption{3D reconstruction result. The red mesh is the target (available from synthetic data), the light blue is the template, and the dark blue is our S2MNet prediction. In the low-MSE case, prediction and target overlap almost perfectly (appearing purple). Even in the high-MSE case, alignment remains strong with only slight visible discrepancies. }\label{result3d}
\end{figure*}

We visualize our result in Figure \ref{result3d}. To demonstrate the robustness of our method, we present two cases: one with the lowest MSE and one with the highest MSE. Each case is shown from four different views. The target 3D representations, depicted in red, were available due to the synthetic nature of our dataset. The light blue mesh represents the template employed in the deformation process, contrasting with the dark blue mesh, which signifies the predicted 3D structure of our S2MNet. There is a clear difference between the template and the prediction mesh. For the case with the lowest MSE ($8.56 \times 10^{-3}$), the prediction closely matches the target, resulting in a perfect overlap that appears purple. Even in the case with the highest MSE ($6.25 \times 10^{-2}$), the prediction and target still align well, with some visible red margins indicating misalignment, but they remain largely similar.

\subsection{Segmentation Overlay Confirms Reconstruction Accuracy}\label{sec:segmentation}

\begin{figure}[h]
\centering
\includegraphics[width= 0.6\linewidth]{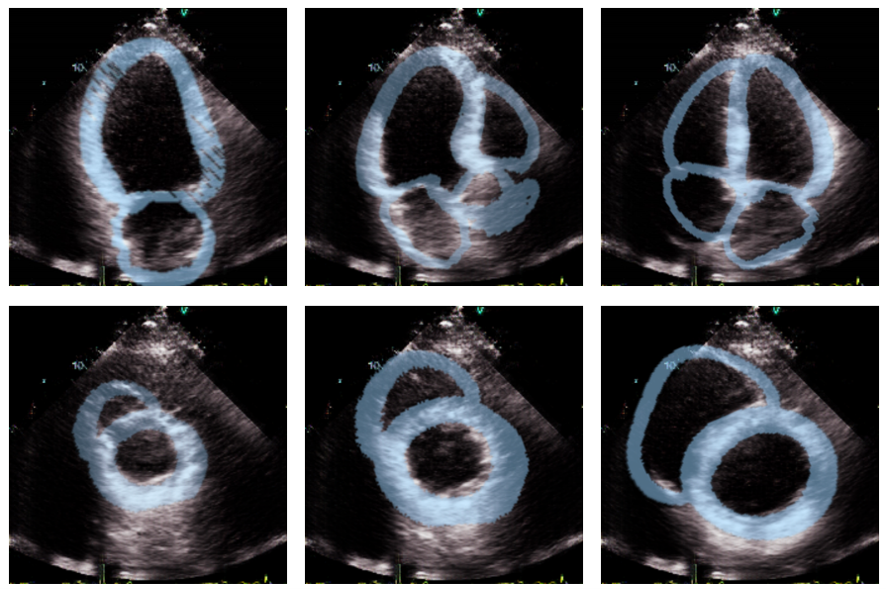}
\caption{Segmentation result: the light blue area represents the segmentation mask, which is obtained by slicing our predicted mesh along the predicted slicing angle. The mask aligns perfectly with the tissue in the echocardiogram images.}\label{seg}
\end{figure}

To further evaluate the accuracy of 3D reconstruction and ensure that it successfully reconstructs all necessary tissues of the cardiac, we slice the reconstructed 3D mesh and overlap it on the input image. Since the angle and coordinates of the input image are unknown. We use a neural network to predict it. The neural network is constructed by ResNet50 \cite{resnet} followed a two branch three fully connect layer, one branch is used to predict angle while the other is used to predict coordinates. The results of our segmentation evaluation approach are shown in Figure \ref{seg}. The segmentation masks which are shown in light blue are able to cover all tissues, which are shown as white pixels in the echocardiogram image.

\subsection{Clinical Analysis on Real Dataset}\label{sec:real}
Ejection fraction (EF) is a key metric for evaluating heart failure. An EF below 50\% may indicate a cardiac condition.
It measures the percentage of blood ejected from the ventricle during each heartbeat, reflecting the heart’s pumping efficiency. It is calculated as $EF(\%) = \frac{SV}{EDV}\times100$, where stroke volume (SV) is the amount of blood pumped out during contraction. Stroke volume itself is defined as the difference between the end-diastolic volume (EDV) and the end-systolic volume (ESV), expressed as $SV=EDV-ESV$. 
% Stroke volume is calculated as the difference between the end-diastolic volume (EDV)—the volume of blood in the ventricle before contraction—and the end-systolic volume (ESV)—the volume remaining after contraction, as shown in Equation \ref{eq::sv}. EF is then computed by dividing the stroke volume by the EDV.

The 3D model generated by S2MNet facilitates efficient left ventricular volume calculation, enabling EF computation. To enhance efficiency, we first apply a $0.1 \times$ subsampling to all points. We then extract the left ventricle region. Our 3D reconstruction method ensures a one-to-one mapping between each point in the prediction and the template, allowing precise region labeling. Finally, we employ the Delaunay triangulation algorithm to generate a mesh for left ventricular volume calculation.

The Pearson correlation coefficient (PCC) \cite{pearson1895note} measures the linear relationship between two data. We use this metric to evaluate the correlation between EF and GLPS,  which is provided by clinicians and serves as the reference standard in this study. The predicted EF and golden standard GLPS for each patient are shown in Table \ref{tab:EF_GLPS}. The PCC between them is -0.82, indicating a strong negative correlation and a strong linear relationship. This result validates the accuracy of our constructed mesh, confirming its reliability for patient diagnosis.

% \begin{equation}\label{eq:PCC}
%     \rho_{X,Y}=\frac{\mathbb{E}[\left.XY\right]-\mathbb{E}[\left.X\right]\mathbb{E}[Y]}{\sqrt{\mathbb{E}[\left.X^2\right.]-\left(\mathbb{E}[\left.X\right]\right)^2}\sqrt{\mathbb{E}[\left.Y^2\right.]-\left(\mathbb{E}[Y]\right)^2}}
% \end{equation}

\begin{table}[]
    \centering
    \begin{tabular}{ c c|c c}
    \hline
          GLPS & EF&GLPS& EF\\
        \hline
        -12.1 & 0.2641&  -14.7 & 0.3166\\
         -12.4 & 0.2859 &-16.7 & 0.3206\\
        -14.9 & 0.2970 & -14.2 & 0.3297\\
         -15.3 & 0.3028& -15.4 & 0.3352 \\
         -14.0 & 0.3035 & -16.8 & 0.3723\\
        -15.8 & 0.3127& -18.2 & 0.3742\\
         -12.6 & 0.3151 &  -20.1 & 0.3861\\ 
\hline
    \end{tabular}
    \caption{GLPS values (provided as the clinical reference standard) and predicted EF values from our reconstructed heart model for individual patients. The strong negative Pearson correlation coefficient (PCC = -0.82) between EF and GLPS demonstrates a robust linear relationship, supporting the reliability of our 3D mesh reconstruction for cardiac function assessment.}
    \label{tab:EF_GLPS}
\end{table}

% \begin{table}[]
%     \centering
%     \begin{tabular}{c c c}
%         Patient ID &  GLPS & EF\\
%         \hline
%         572118  & -12.1 & 0.2641\\
%         534476  & -12.4 & 0.2859\\
%         568243  & -14.9 & 0.2970 \\
%         568863  & -15.3 & 0.3028 \\
%         568877  & -14.0 & 0.3035\\
%         570477  & -15.8 & 0.3127\\
%         571202  & -12.6 & 0.3151 \\
%         571932  & -14.7 & 0.3166\\
%         572103  & -16.7 & 0.3206\\
%         572380  & -14.2 & 0.3297 \\
%         572738  & -15.4 & 0.3352\\
%         572957  & -16.8 & 0.3723\\
%         903674  & -18.2 & 0.3742 \\
%         569820  & -20.1 & 0.3861\\  
% \hline
%     \end{tabular}
%     \caption{GLPS values (provided as the clinical reference standard) and predicted EF values from our reconstructed heart model for individual patients. The strong negative Pearson correlation coefficient (PCC = -0.82) between EF and GLPS demonstrates a robust linear relationship, supporting the reliability of our 3D mesh reconstruction for cardiac function assessment.}
%     \label{tab:EF_GLPS}
% \end{table}

\section{Discussion}\label{sec:discussion}

In this study, we proposed a novel framework for reconstructing high-quality 3D cardiac models from 2D echocardiogram images. Our approach integrates synthetic data generation, image style transfer using CycleGAN, deformation field computation, and 3D reconstruction with S2MNet. By leveraging predefined echocardiogram views and style-adapted synthetic training data, our method achieves superior anatomical continuity and spatial coherence compared to conventional reconstruction techniques.

%The experimental results demonstrate that our approach significantly enhances the fidelity of 3D cardiac reconstructions. The CycleGAN-transformed images closely resemble real echocardiogram images, while our S2MNet effectively captures intricate cardiac structures with minimal artifacts. Quantitative evaluations indicate that our method surpasses existing voxel-based approaches in IoU accuracy, achieving a high level of structural consistency with target 3D models. Moreover, the reconstruction results align well with the input echocardiogram images, and the EF computed from our mesh corresponds with the GLPS provided by doctors. This confirms the robustness of the model in different perspectives and patient anatomies, validating its applicability for medical diagnosis.

Moving forward, we aim to refine our approach by integrating uncertainty quantification to assess confidence in the reconstructed models. Furthermore, incorporating real-time inference capabilities could make this method more applicable in clinical settings, supporting real-time 3D cardiac visualization from live echocardiogram scans. Finally, validating our framework on diverse patient datasets, including pathological cases, will be essential for ensuring its robustness and clinical reliability.

% ==============
% # REFERENCES #
% ==============

{
    \small
    \bibliographystyle{ieeenat_fullname}
    \bibliography{ref}

\begin{thebibliography}{44}
\providecommand{\natexlab}[1]{#1}
\providecommand{\url}[1]{\texttt{#1}}
\expandafter\ifx\csname urlstyle\endcsname\relax
  \providecommand{\doi}[1]{doi: #1}\else
  \providecommand{\doi}{doi: \begingroup \urlstyle{rm}\Url}\fi

\bibitem[Canny(1986)]{canny1986computational}
John Canny.
\newblock A computational approach to edge detection.
\newblock \emph{IEEE Transactions on pattern analysis and machine intelligence}, \penalty0 (6):\penalty0 679--698, 1986.

\bibitem[Chen et~al.(2023)Chen, Wang, Yuan, Yang, and Yue]{PCR1}
Guangyan Chen, Meiling Wang, Li Yuan, Yi Yang, and Yufeng Yue.
\newblock Rethinking point cloud registration as masking and reconstruction.
\newblock In \emph{Proceedings of the IEEE/CVF International Conference on Computer Vision}, pages 17717--17727, 2023.

\bibitem[Chen et~al.(2019)Chen, Han, Xu, and Su]{msv2}
Rui Chen, Songfang Han, Jing Xu, and Hao Su.
\newblock Point-based multi-view stereo network.
\newblock In \emph{Proceedings of the IEEE/CVF international conference on computer vision}, pages 1538--1547, 2019.

\bibitem[Everingham et~al.(2010)Everingham, Van~Gool, Williams, Winn, and Zisserman]{iou}
Mark Everingham, Luc Van~Gool, Christopher Williams, John Winn, and Andrew Zisserman.
\newblock The pascal visual object classes (voc) challenge.
\newblock \emph{International Journal of Computer Vision}, 88:\penalty0 303--338, 2010.

\bibitem[Fan et~al.(2017)Fan, Su, and Guibas]{pointoutnet}
Haoqiang Fan, Hao Su, and Leonidas~J Guibas.
\newblock A point set generation network for 3d object reconstruction from a single image.
\newblock In \emph{Proceedings of the IEEE conference on computer vision and pattern recognition}, pages 605--613, 2017.

\bibitem[Foley et~al.(1990)Foley, van Dam, Feiner, and Hughes]{3dgraph}
James~D. Foley, Andries van Dam, Steven~K. Feiner, and John~F. Hughes.
\newblock \emph{Computer graphics: principles and practice (2nd ed.)}.
\newblock Addison-Wesley Longman Publishing Co., Inc., USA, 1990.

\bibitem[Gao et~al.(2022)Gao, Gao, He, Lu, Xu, and Li]{nerfreview}
Kyle Gao, Yina Gao, Hongjie He, Dening Lu, Linlin Xu, and Jonathan Li.
\newblock Nerf: Neural radiance field in 3d vision, a comprehensive review.
\newblock \emph{arXiv preprint arXiv:2210.00379}, 2022.

\bibitem[Gillam and Marcoff(2024)]{gillam2024echocardiography}
Linda~D. Gillam and Lauren Marcoff.
\newblock Echocardiography: Past, present, and future.
\newblock \emph{Circulation: Cardiovascular Imaging}, 17\penalty0 (4):\penalty0 e016517, 2024.

\bibitem[Goesele et~al.(2006)Goesele, Curless, and Seitz]{msv1}
Michael Goesele, Brian Curless, and Steven~M Seitz.
\newblock Multi-view stereo revisited.
\newblock In \emph{2006 IEEE Computer Society Conference on Computer Vision and Pattern Recognition (CVPR'06)}, pages 2402--2409. IEEE, 2006.

\bibitem[Hartley and Zisserman(2004)]{hartley2004multiple}
Richard Hartley and Andrew Zisserman.
\newblock \emph{Multiple View Geometry in Computer Vision}.
\newblock Cambridge University Press, 2004.

\bibitem[He et~al.(2016)He, Zhang, Ren, and Sun]{resnet}
Kaiming He, Xiangyu Zhang, Shaoqing Ren, and Jian Sun.
\newblock Deep residual learning for image recognition.
\newblock In \emph{Proceedings of the IEEE conference on computer vision and pattern recognition}, pages 770--778, 2016.

\bibitem[Houck et~al.(2006)Houck, Cooke, and Gill]{Houck2006Live3E}
Robin~C Houck, Jason Cooke, and Edward~A. Gill.
\newblock Live 3d echocardiography: a replacement for traditional 2d echocardiography?
\newblock \emph{AJR. American journal of roentgenology}, 187 4:\penalty0 1092--106, 2006.

\bibitem[Hu et~al.(2023)Hu, Huang, Li, Gao, Yin, Qi, Wu, Chen, Ma, Shi, Li, Maus, Huang, Lu, Lin, Zhou, Lou, Gu, Chen, and Xu]{nature1}
Hongjie Hu, Hao Huang, Mohan Li, Xiaoxiang Gao, Lu Yin, Ruixiang Qi, Cervantes Wu, Xiangjun Chen, Yuxiang Ma, Keren Shi, Chenghai Li, Timothy Maus, Brady Huang, Chengchangfeng Lu, Muyang Lin, Sai Zhou, Zhiyuan Lou, Yue Gu, Yimu Chen, and Sheng Xu.
\newblock A wearable cardiac ultrasound imager.
\newblock \emph{Nature}, 613:\penalty0 667--675, 2023.

\bibitem[Izadi et~al.(2011)Izadi, Kim, Hilliges, Molyneaux, Newcombe, Kohli, Shotton, Hodges, Freeman, Davison, et~al.]{KinectFusion1}
Shahram Izadi, David Kim, Otmar Hilliges, David Molyneaux, Richard Newcombe, Pushmeet Kohli, Jamie Shotton, Steve Hodges, Dustin Freeman, Andrew Davison, et~al.
\newblock Kinectfusion: real-time 3d reconstruction and interaction using a moving depth camera.
\newblock In \emph{Proceedings of the 24th annual ACM symposium on User interface software and technology}, pages 559--568, 2011.

\bibitem[Jin et~al.(2005)Jin, Soatto, and Yezzi]{msv3}
Hailin Jin, Stefano Soatto, and Anthony~J Yezzi.
\newblock Multi-view stereo reconstruction of dense shape and complex appearance.
\newblock \emph{International Journal of Computer Vision}, 63:\penalty0 175--189, 2005.

\bibitem[Jin et~al.(2020)Jin, Jiang, and Cai]{dl3dsurvey}
Yiwei Jin, Diqiong Jiang, and Ming Cai.
\newblock 3d reconstruction using deep learning: a survey.
\newblock \emph{Communications in Information and Systems}, 20\penalty0 (4):\penalty0 389--413, 2020.

\bibitem[Lengyel(2011)]{lengyel2011mathematics}
Eric Lengyel.
\newblock \emph{Mathematics for 3D Game Programming and Computer Graphics}.
\newblock Cengage Learning, 3rd edition, 2011.

\bibitem[Maresca et~al.(2017)Maresca, Correia, Villemain, Alain, Sambin, Ghaleh, and Pernot]{article}
David Maresca, Mafalda Correia, Olivier Villemain, Bize Alain, Lucien Sambin, Bijan Ghaleh, and Mathieu Pernot.
\newblock Noninvasive imaging of the coronary vasculature using ultrafast ultrasound.
\newblock \emph{JACC: Cardiovascular Imaging}, 11, 2017.

\bibitem[Mildenhall et~al.(2021)Mildenhall, Srinivasan, Tancik, Barron, Ramamoorthi, and Ng]{nerf1}
Ben Mildenhall, Pratul~P Srinivasan, Matthew Tancik, Jonathan~T Barron, Ravi Ramamoorthi, and Ren Ng.
\newblock Nerf: Representing scenes as neural radiance fields for view synthesis.
\newblock \emph{Communications of the ACM}, 65\penalty0 (1):\penalty0 99--106, 2021.

\bibitem[Moncef and M'hamed(2020)]{2dto3d}
Aharchi Moncef and Aït~Kbir M'hamed.
\newblock A review on 3d reconstruction techniques from 2d images.
\newblock pages 510--522, 2020.

\bibitem[Navaneet et~al.(2020)Navaneet, Mathew, Kashyap, Hung, Jampani, and Babu]{encoder2}
KL Navaneet, Ansu Mathew, Shashank Kashyap, Wei-Chih Hung, Varun Jampani, and R~Venkatesh Babu.
\newblock From image collections to point clouds with self-supervised shape and pose networks.
\newblock In \emph{Proceedings of the IEEE/CVF Conference on Computer Vision and Pattern Recognition}, pages 1132--1140, 2020.

\bibitem[Newcombe et~al.(2011)Newcombe, Izadi, Hilliges, Molyneaux, Kim, Davison, Kohi, Shotton, Hodges, and Fitzgibbon]{KinectFusion2}
Richard~A. Newcombe, Shahram Izadi, Otmar Hilliges, David Molyneaux, David Kim, Andrew~J. Davison, Pushmeet Kohi, Jamie Shotton, Steve Hodges, and Andrew Fitzgibbon.
\newblock Kinectfusion: Real-time dense surface mapping and tracking.
\newblock In \emph{2011 10th IEEE International Symposium on Mixed and Augmented Reality}, pages 127--136, 2011.

\bibitem[Oliensis(2000)]{SFM1}
John Oliensis.
\newblock A critique of structure-from-motion algorithms.
\newblock \emph{Computer Vision and Image Understanding}, 80\penalty0 (2):\penalty0 172--214, 2000.

\bibitem[Pearson(1895)]{pearson1895note}
Karl Pearson.
\newblock Note on regression and inheritance in the case of two parents.
\newblock \emph{Proceedings of the Royal Society of London}, 58:\penalty0 240--242, 1895.

\bibitem[Peng et~al.(2023)Peng, Lin, Wu, and Cao]{PCR3}
Yeping Peng, Shengdong Lin, Hongkun Wu, and Guangzhong Cao.
\newblock Point cloud registration based on fast point feature histogram descriptors for 3d reconstruction of trees.
\newblock \emph{Remote Sensing}, 15\penalty0 (15):\penalty0 3775, 2023.

\bibitem[Pishchulin et~al.(2017)Pishchulin, Wuhrer, Helten, Theobalt, and Schiele]{bodymesh}
Leonid Pishchulin, Stefanie Wuhrer, Thomas Helten, Christian Theobalt, and Bernt Schiele.
\newblock Building statistical shape spaces for 3d human modeling.
\newblock \emph{Pattern Recognition}, 2017.

\bibitem[Popov et~al.(2020)Popov, Bauszat, and Ferrari]{popov2020corenet}
Stefan Popov, Pablo Bauszat, and Vittorio Ferrari.
\newblock Corenet: Coherent 3d scene reconstruction from a single rgb image.
\newblock In \emph{Computer Vision--ECCV 2020: 16th European Conference, Glasgow, UK, August 23--28, 2020, Proceedings, Part II 16}, pages 366--383. Springer, 2020.

\bibitem[Pumarola et~al.(2021)Pumarola, Corona, Pons-Moll, and Moreno-Noguer]{dnerf}
Albert Pumarola, Enric Corona, Gerard Pons-Moll, and Francesc Moreno-Noguer.
\newblock D-nerf: Neural radiance fields for dynamic scenes.
\newblock In \emph{Proceedings of the IEEE/CVF Conference on Computer Vision and Pattern Recognition}, pages 10318--10327, 2021.

\bibitem[Rodero et~al.(2021)Rodero, Strocchi, Marciniak, Longobardi, Whitaker, O'Neill, Gillette, Augustin, Plank, Vigmond, Lamata, and NIederer]{3Ddataset}
Cristobal Rodero, Marina Strocchi, Maciej Marciniak, Stefano Longobardi, John Whitaker, Mark~D. O'Neill, Karli Gillette, Christoph Augustin, Gernot Plank, Edward~J. Vigmond, Pablo Lamata, and Steven NIederer.
\newblock {Virtual cohort of 1000 synthetic heart meshes from adult human healthy population}, 2021.

\bibitem[Shen et~al.(2021)Shen, Feydy, Liu, Curiale, San Jose~Estepar, San Jose~Estepar, and Niethammer]{nipsot}
Zhengyang Shen, Jean Feydy, Peirong Liu, Ariel~H Curiale, Ruben San Jose~Estepar, Raul San Jose~Estepar, and Marc Niethammer.
\newblock Accurate point cloud registration with robust optimal transport.
\newblock \emph{Advances in Neural Information Processing Systems}, 34:\penalty0 5373--5389, 2021.

\bibitem[Stojanovski et~al.(2022)Stojanovski, Hermida, Muffoletto, Lamata, Beqiri, and Gomez]{e-p2vforheart}
David Stojanovski, Uxio Hermida, Marica Muffoletto, Pablo Lamata, Arian Beqiri, and Alberto Gomez.
\newblock Efficient pix2vox++ for 3d cardiac reconstruction from 2d echo views.
\newblock In \emph{Simplifying Medical Ultrasound}, pages 86--95, Cham, 2022. Springer International Publishing.

\bibitem[Turton and Ender(2017)]{turton2017role}
E.~W. Turton and J. Ender.
\newblock Role of 3d echocardiography in cardiac surgery: Strengths and limitations.
\newblock \emph{Current Anesthesiology Reports}, 7\penalty0 (3):\penalty0 291--298, 2017.

\bibitem[Wu and Takeuchi(2017)]{2dlimit}
Victor Chien-Chia Wu and Masaaki Takeuchi.
\newblock Three-dimensional echocardiography: Current status and real-life applications.
\newblock \emph{Acta Cardiologica Sinica}, 33:\penalty0 107--118, 2017.

\bibitem[Wu et~al.(2019)Wu, Wang, Li, Liu, and Fuxin]{wu2019pointpwc}
Wenxuan Wu, Zhiyuan Wang, Zhuwen Li, Wei Liu, and Li Fuxin.
\newblock Pointpwc-net: A coarse-to-fine network for supervised and self-supervised scene flow estimation on 3d point clouds.
\newblock \emph{arXiv preprint arXiv:1911.12408}, 2019.

\bibitem[Xie et~al.(2019)Xie, Yao, Sun, Zhou, and Zhang]{pix2vox}
Haozhe Xie, Hongxun Yao, Xiaoshuai Sun, Shangchen Zhou, and Shengping Zhang.
\newblock Pix2vox: Context-aware 3d reconstruction from single and multi-view images.
\newblock In \emph{ICCV}, 2019.

\bibitem[Yan et~al.(2016)Yan, Yang, Yumer, Guo, and Lee]{encoder1}
Xinchen Yan, Jimei Yang, Ersin Yumer, Yijie Guo, and Honglak Lee.
\newblock Perspective transformer nets: Learning single-view 3d object reconstruction without 3d supervision.
\newblock \emph{Advances in neural information processing systems}, 29, 2016.

\bibitem[Yang et~al.(2013)Yang, Daimon, Ishii, Kawata, Miyazaki, Hirose, Ichikawa, Chiang, Suzuki, Miyauchi, and Daida]{RN11}
B. Yang, M. Daimon, K. Ishii, T. Kawata, S. Miyazaki, K. Hirose, R. Ichikawa, S.~J. Chiang, H. Suzuki, K. Miyauchi, and H. Daida.
\newblock Prediction of coronary artery stenosis at rest in patients with normal left ventricular wall motion. segmental analyses using strain imaging diastolic index.
\newblock \emph{Int Heart J}, 54\penalty0 (5):\penalty0 266--72, 2013.

\bibitem[Yang et~al.(2020)Yang, Shi, and Carlone]{PCR2}
Heng Yang, Jingnan Shi, and Luca Carlone.
\newblock Teaser: Fast and certifiable point cloud registration.
\newblock \emph{IEEE Transactions on Robotics}, 37\penalty0 (2):\penalty0 314--333, 2020.

\bibitem[Yeom et~al.(2014)Yeom, Nam, Jin, Paeng, and Lee]{3Drebifur}
Eunseop Yeom, Kweon-Ho Nam, Changzhu Jin, Dong-Guk Paeng, and Sang-Joon Lee.
\newblock 3d reconstruction of a carotid bifurcation from 2d transversal ultrasound images.
\newblock \emph{Ultrasonics}, 54\penalty0 (8):\penalty0 2184--2192, 2014.

\bibitem[Zhang et~al.(2021)Zhang, Zhu, Chen, Yang, Cheng, Li, Zhong, and Wang]{zhangjecho}
Jingyi Zhang, Huolan Zhu, Yongkai Chen, Chenguang Yang, Huimin Cheng, Yi Li, Wenxuan Zhong, and Fang Wang.
\newblock Ensemble machine learning approach for screening of coronary heart disease based on echocardiography and risk factors.
\newblock \emph{BMC Medical Informatics and Decision Making}, 21, 2021.

\bibitem[Zhang et~al.(2020)Zhang, Riegler, Snavely, and Koltun]{nerf++}
Kai Zhang, Gernot Riegler, Noah Snavely, and Vladlen Koltun.
\newblock Nerf++: Analyzing and improving neural radiance fields.
\newblock \emph{arXiv preprint arXiv:2010.07492}, 2020.

\bibitem[Zhu et~al.(2020)Zhu, Park, Isola, and Efros]{cyclegan}
Jun-Yan Zhu, Taesung Park, Phillip Isola, and Alexei~A. Efros.
\newblock Unpaired image-to-image translation using cycle-consistent adversarial networks, 2020.

\bibitem[Zou et~al.(2023)Zou, Huang, Gao, Zhang, Wang, and Wan]{3Dresnet}
Qin Zou, Yuqing Huang, Junling Gao, Bo Zhang, Diya Wang, and Mingxi Wan.
\newblock Three-dimensional ultrasound image reconstruction based on 3d-resnet in the musculoskeletal system using a 1d probe: ex vivo and in vivo feasibility studies.
\newblock \emph{Physics in Medicine \& Biology}, 68\penalty0 (16):\penalty0 165003, 2023.

\bibitem[Özyeşil et~al.(2017)Özyeşil, Voroninski, Basri, and Singer]{SFM2}
Onur Özyeşil, Vladislav Voroninski, Ronen Basri, and Amit Singer.
\newblock A survey of structure from motion.
\newblock \emph{Acta Numerica}, 26:\penalty0 305–364, 2017.

\end{thebibliography}
}

\end{document}